\documentclass[12pt,floats]{article}

\usepackage{mathrsfs}
\usepackage{amsmath}
\usepackage{amssymb}
\usepackage{graphicx}
\usepackage{longtable}

\usepackage{bm}
\usepackage{epsfig}
\usepackage{citesort}

\begin{document}


\title{Mesonic excitations and $\pi$--$\pi$ scattering lengths at finite
temperature in the two-flavor Polyakov--Nambu--Jona-Lasinio model}

\author{{Wei-jie Fu$^{a}$,
and Yu-xin Liu$^{a,b,}$\thanks{Corresponding author, e-mail address: yxliu@pku.edu.cn} }\\[3mm]
\normalsize{$^a$ Department of Physics and State Key Laboratory of
Nuclear Physics and Technology,}\\
\normalsize{ Peking University, Beijing 100871, China}\\
\normalsize{$^b$ Center of Theoretical Nuclear Physics, National
Laboratory of Heavy Ion Accelerator,}\\ \normalsize{ Lanzhou 730000,
China}  }

%
%
%
%

\maketitle

\begin{abstract}
The mesonic excitations and $s$-wave $\pi$--$\pi$ scattering lengths
at finite temperature are studied in the two-flavor
Polyakov--Nambu--Jona-Lasinio (PNJL) model. The masses of $\pi$
meson and $\sigma$ meson, pion-decay constant, the pion-quark
coupling strength, and the scattering lengths $a_{0}$ and $a_{2}$ at
finite temperature are calculated in the PNJL model with two forms
of Polyakov-loop effective potential. The obtained results are
almost independent of the choice of the effective potentials. The
calculated results in the PNJL model are also compared with those in
the conventional Nambu--Jona-Lasinio model and indicate that the
effect of color confinement screens the effect of temperature below
the critical one in the PNJL model. Furthermore, the
Goldberger-Treiman relation and the Gell-Mann--Oakes--Renner
relation are extended to the case at finite temperature in the PNJL
model.
\end{abstract}

\noindent{\bf PACS Numbers: }
{12.38.Aw, 
      11.30.Rd, 
      13.75.Lb, 
      14.40.Aq  
      }


\newpage

\section{Introduction}
\vspace{5pt}

QCD thermodynamics and phase diagram, especially about the
restoration of the chiral symmetry and the deconfinement phase
transition which are expected to occur in ultra-relativistic
heavy-ion
collisions~\cite{Shuryak2004,Gyulassy2005,Shuryak2005,Arsene2005,Back2005,Adams2005,Adcox2005,Blaizot2007}
or in the interior of neutron
stars~\cite{Weber2005,Alford2007,Alford2008,Fu2008b}, has been a
subject of intense investigation in recent years. One significant
aspect to investigate the restoration of the chiral or axial
symmetry and the deconfinement phase transition is to study the
variation of properties of particles propagating in hot and/or dense
medium~\cite{Costa2004,Hansen2007,Costa2008}. In this work, we focus
on the influence of a hot medium on the properties of light
pseudoscalar ($\pi$) and scalar ($\sigma$) mesons,  and $\pi$--$\pi$
scattering lengths. Special attentions are paid to their dramatic
variations near the regime where the chiral phase transition and the
deconfinement phase transition occur. We expect to extract the
signals of phase transition from mesonic excitations and
$\pi$--$\pi$ interactions in the hot medium.

A promising phenomenological approach to study the low-energy
processes involving the pseudoscalar and scalar mesons at zero
temperature and finite temperature is the Nambu--Jona-Lasinio (NJL)
model~\cite{Nambu1961,Volkov1984,Klevansky1992,Hatsuda1994,Alkofer1996,Buballa2005}.
The most important advantage of the NJL model is that it introduces
a mechanism of the dynamical breaking of chiral symmetry (due to the
quark-antiquark condensate). However, the NJL model has its
disadvantage, which is the lack of the description of color
confinement. To include some effects of color confinement, a
Polyakov-loop improved Nambu--Jona-Lasinio (PNJL) model has been
developed recent
years~\cite{Meisinger9602,Pisarski2000,Fukushima2004,Mocsy2004,Arriola2006,Ratti2006a,Ratti2006b,Fu2008,Ciminale2008}.
In the PNJL model, the Polyakov-loop as a classical field couples to
quarks and thus suppresses the contributions from wrong degrees of
freedom (color non-singlet) to the thermodynamics below the critical
temperature. Therefore, the introduction of the Polyakov-loop
represents some aspects of the color confinement, at least on the
level of statistics~\cite{Fu2008}. The validity of the PNJL model
has been confirmed in a series of works by confronting the PNJL
results with the lattice QCD
data~\cite{Ratti2006a,Ratti2006b,Ghosh2006,Ratti2006c,zhang2006}.
The phase structure and thermodynamics in the PNJL model have
recently been explored
extensively~\cite{Sasaki2006,Weise2007,Fu2008,Ciminale2008,Ghosh2008,Abuki2008,Tuominen2008,Abuki2008b,Kashiwa2008,Costa2008b,Fukushima2008a,Contrera2008,Fukushima2008b,Mukherjee2007,Hiller2008},
and the impact of Polyakov-loop dynamics on the chiral
susceptibility or quark number
susceptibility~\cite{Sasaki2006,Weise2007}, QCD critical
endpoint~\cite{Kashiwa2008,Costa2008b} and critical
surface~\cite{Fukushima2008a}, and the color superconductivity phase
transition~\cite{Ratti2006b,Ciminale2007,Abuki2008c,Dumm2008} have
attracted lots of interests. Furthermore, fluctuations beyond the
mean field approximation have been included in the PNJL
model~\cite{Blaschke2007,Hell2007}, and the PNJL model has also been
extended to the regime of imaginary chemical
potential~\cite{Sakai2008,Sakai2008b,Kashiwa2008b,Sakai2008c} and
$0+1$ dimensions~\cite{Dusling2008}, and applied to analyze the
flavors of quark-gluon-plasma~\cite{Mueller2008} and the isentropic
trajectories on QCD phase diagram~\cite{Fukushima2009}.

The properties of pseudoscalar and scalar mesons at finite
temperature for two~\cite{Hansen2007} and three~\cite{Costa2008}
flavor systems have also been investigated in the PNJL model. In
Ref.~\cite{Hansen2007}, The mesonic correlators and spectral
functions for $\pi$ and $\sigma$ mesons were obtained. It was found
that the $\pi$-$\sigma$ degeneracy in the chiral symmetry restored
phase was still satisfied after coupling quarks to the Polyakov-loop
and the role of $\pi$ meson as Goldstone boson was also confirmed in
the PNJL model. It was also found that, although the PNJL model can
not cure the problem of the conventional NJL model as for the
unphysical width of the $\sigma$ meson, the PNJL results on the
decay width improved slightly the NJL ones~\cite{Hansen2007}. In
order to further study the broken chiral symmetry and its
restoration in the mesonic sector in the PNJL model which makes the
investigation of the interplay between the restoration of chiral
symmetry and the deconfinement phase transition possible, it is
necessary to study the Goldberger-Treiman
relation~\cite{Goldberger1958} and the Gell-Mann--Oakes--Renner
relation~\cite{Gell1968} which are direct results due to the chiral
symmetry breaking. Furthermore, one of the most fundamental hadronic
processes of QCD at the mesonic level, the pion-pion scattering,
$\pi+\pi\rightarrow\pi+\pi$,  at finite temperature, which provides
a direct link between the theoretical formalism of chiral symmetry
and experiment, also deserves to be investigated. In this work, we
will then study the problems mentioned above in the PNJL model.

The paper is organized as follows. In Sec. II we simply review the
formalism of the two flavor PNJL model. In Sec. III we discuss the
mesonic excitations at finite temperature in the PNJL model. The
dependence of the pion-decay constant, pion-quark coupling strength,
and the relation between the mass of $\sigma$ meson and that of
$\pi$ meson on the temperature are studied in the PNJL model. We
also extend the Goldberger-Treiman relation and
Gell-Mann--Oakes--Renner relation to a formalism which is
appropriate at finite temperature. In Sec. IV we study the $s$-wave
$\pi$--$\pi$ scattering lengths in the PNJL model and compare the
results in the PNJL model with those in the conventional NJL.
Finally, in Sec. V, we give a summary and conclusions.

\section{The PNJL model}

The Lagrangian density for the two-flavor PNJL model is given
as~\cite{Ratti2006a}
\begin{eqnarray}
\mathcal{L}_{PNJL}&=&\bar{\psi}\left(i\gamma_{\mu}D^{\mu}-\hat{m}_{0}\right)\psi
 +G\left[\left(\bar{\psi}\psi\right)^{2}
 +\left(\bar{\psi}i\gamma_{5}\vec{\tau}\psi\right)^{2}\right] \nonumber \\
&& -\mathcal{U}\left(\Phi[A],\bar{\Phi}[A] \,
,T\right),\label{lagragian}
\end{eqnarray}
where $\psi=(\psi_{u},\psi_{d})^{T}$ is the quark field,
\begin{equation}
D^{\mu}=\partial^{\mu}-iA^{\mu}\quad\textrm{with}\quad
A^{\mu}=\delta^{u}_{0}A^{0}\quad\textrm{,}\quad
A^{0}=g\mathcal{A}^{0}_{a}\frac{\lambda_{a}}{2}=-iA_4.
\end{equation}
The gauge coupling $g$ is combined with the SU(3) gauge field
$\mathcal{A}^{\mu}_{a}(x)$ to define $A^{\mu}(x)$ for convenience
and $\lambda_{a}$ are the Gell-Mann matrices in color space.
$\hat{m}_{0}=\textrm{diag}(m_{u},m_{d})$ is the current quark mass
matrix. Throughout this work, we take $m_{u}=m_{d}\equiv m_{0}$,
assuming the isospin symmetry is reserved on the Lagrangian level.
The four-fermion interaction with an effective coupling strength $G$
for scalar and pseudoscalar channels has $\mathrm{SU_{V}}(2)\times
\mathrm{SU_{A}}(2)\times \mathrm{U_{V}}(1)$ symmetry, which is
broken to $\mathrm{SU_{V}}(2)\times \mathrm{U_{V}}(1)$ when
$m_{0}\neq 0$. Here $\tau^{a}(a=1,2,3)$ in the Lagrangian density
(Eq.~\eqref{lagragian}) are Pauli matrices in flavor space.

The $\mathcal{U}\left(\Phi,\bar{\Phi},T\right)$ in the Lagrangian
density is the Polyakov-loop effective potential, which controls the
Polyakov-loop dynamics and can be expressed in terms of the trace of
the Polyakov-loop $\Phi=(\mathrm{Tr}_{c}L)/N_{c}$ and its conjugate
$\bar{\Phi}=(\mathrm{Tr}_{c}L^{\dag})/N_{c}$. Here the Polyakov-loop
$L$ is a matrix in color space, which can be explicitly given
as~\cite{Ratti2006a}
\begin{equation}
L\left(\vec{x}\right)=\mathcal{P}\exp\left[i\int_{0}^{\beta}d\tau\,
A_{4}\left(\vec{x},\tau\right)\right]
                     =\exp\left[i \beta A_{4} \right]\, ,
\end{equation}
where $\beta=1/T$ is the inverse of the temperature. The
Polyakov-loop effective potential has the $Z(3)$ center symmetry
like the pure-gauge QCD Lagrangian. When the temperature is lower
than a critical value ($T_{0}\simeq 270\,\mathrm{MeV}$ in pure gauge
QCD~\cite{Ratti2006a}), the value of $\Phi$ (and $\bar{\Phi}$) which
minimizes the Polyakov-loop effective potential is zero, meaning
that the phase is color confined and has the $Z(3)$ symmetry.
However, when the temperature is above the critical temperature
$T_{0}$, $\Phi$ develops a nonzero value which minimizes the
effective potential and the system is transited from a $Z(3)$
symmetric, confined phase to a $Z(3)$ symmetry broken, deconfined
phase. The temperature dependent Polyakov-loop effective potential
is chosen to reproduce the lattice data for both the expectation
value of the Polyakov-loop~\cite{Kaczmarek2002} and some
thermodynamic quantities~\cite{Boyd1996}. In the PNJL Lagrangian in
Eq.~\eqref{lagragian}, the coupling between the Polyakov-loop and
quarks is uniquely determined by the covariant derivative $D_{\mu}$.

In previous works, two possible forms for the Polyakov-loop
effective potential have been well developed. Following our previous
work~\cite{Fu2008}, we denote them as
$\mathcal{U}_{\mathrm{pol}}(\Phi,\bar{\Phi},T)$ and
$\mathcal{U}_{\mathrm{imp}}(\Phi,\bar{\Phi},T)$, respectively. The
former is a polynomial in $\Phi$ and $\bar{\Phi}$~\cite{Ratti2006a}
and the latter is an improved effective potential in which the
higher order polynomial terms in $\Phi$ and $\bar{\Phi}$ are
replaced by a logarithm~\cite{Ratti2006b}. Both the effective
potentials are taken in our work to investigate whether our results
depend on the details of the Polyakov-loop effective potential.
These two effective potentials have the following forms
\begin{equation}
\frac{\mathcal{U}_{\mathrm{pol}}\left(\Phi,\bar{\Phi},T\right)}{T^{4}}
= -\frac{b_{2}(T)}{2}\bar{\Phi}\Phi -\frac{b_{3}}{6}
(\Phi^{3}+{\bar{\Phi}}^{3})+\frac{b_{4}}{4}(\bar{\Phi}\Phi)^{2} \, ,
\end{equation}
with
\begin{equation}
b_{2}(T)=a_{0}+a_{1}\left(\frac{T_{0}}{T}\right)+a_{2}
{\left(\frac{T_{0}}{T}\right)}^{2}
+a_{3}{\left(\frac{T_{0}}{T}\right)}^{3},
\end{equation}
and
\begin{eqnarray}
\frac{\mathcal{U}_{\mathrm{imp}}\left(\Phi,\bar{\Phi},T\right)}{T^{4}}
& = &-\frac{1}{2}A(T)\bar{\Phi}\Phi
+B(T)\ln\left[1-6\bar{\Phi}\Phi+4({\bar{\Phi}}^{3}+\Phi^{3})
-3(\bar{\Phi}\Phi)^{2}\right]  \, ,
\end{eqnarray}
with
\begin{equation}
A(T)=A_{0}+A_{1}\left(\frac{T_{0}}{T}\right)
     +A_{2}{\left(\frac{T_{0}}{T}\right)}^{2},\quad
B(T)=B_{3}{\left(\frac{T_{0}}{T}\right)}^{3}.
\end{equation}
A precise fit of the parameters in these two effective potentials
has been performed to reproduce some pure-gauge lattice QCD data in
Refs.~\cite{Ratti2006a,Ratti2006b}. The results are listed in
Table~\ref{pol_para}, Table~\ref{imp_para}, respectively. The
parameter $T_{0}$ is the critical temperature for the deconfinement
phase transition to take place in the pure-gauge QCD and $T_{0}$ is
chosen to be $270\,\mathrm{MeV}$ according to the lattice
calculations.

\begin{table}[htb]
\begin{center}
\caption{Parameters for the polynomial effective potential
$\mathcal{U}_{\mathrm{pol}}$} \label{pol_para}
\begin{tabular}{cccccc}
\hline \hline \vspace{0.1cm}
$a_{0}$\qquad\qquad&$a_{1}$\qquad\qquad&$a_{2}$\qquad\qquad&$a_{3}$\qquad\qquad&
$b_{3}$\qquad\qquad& $b_{4}$\\
\hline 6.75 \qquad\qquad& $-1.95$ \qquad\qquad& 2.625\qquad\qquad& $-7.44$\qquad\qquad& 0.75\qquad\qquad& 7.5\\
\hline
\end{tabular}
\end{center}
\end{table}

\begin{table}[htb]
\begin{center}
\caption{Parameters for the improved effective potential
$\mathcal{U}_{\mathrm{imp}}$} \label{imp_para}
\begin{tabular}{cccc}
\hline \hline \vspace{0.1cm}
$A_{0}$\qquad\qquad&$A_{1}$\qquad\qquad&$A_{2}$\qquad\qquad&$B_{3}$\\
\hline 3.51 \qquad\qquad& $-2.47$ \qquad\qquad& 15.2\qquad\qquad& $-1.75$\\
\hline
\end{tabular}
\end{center}
\end{table}

In the NJL sector of the model, three parameters need to be
determined: the three-momentum cutoff $\Lambda$, the current quark
mass $m_{0}$, and the coupling strength $G$. In our work we employ
the zero-temperature values of the quark condensate, pion decay
constant and the mass of pion to fix the parameters. The obtained
results are given in Table~\ref{NJL_para}.

\begin{table}[htb]
\begin{center}
\caption{Parameters of the NJL sector of the model
and the physical quantities being fitted } \label{NJL_para}
\begin{tabular}{cccccc}
\hline \hline \vspace{0.1cm}
$\Lambda\,(\mathrm{MeV})$\qquad\qquad&$G\,({\mathrm{GeV}}^{-2})$\qquad\qquad
&
$m_{0}\,(\mathrm{MeV})$\qquad\qquad&$|\langle\bar{\psi}_{u}\psi_{u}\rangle|^{1/3}\,(\mathrm{MeV})$\qquad\qquad&
$f_{\pi}\,(\mathrm{MeV})$\qquad\qquad& $m_{\pi}\,(\mathrm{MeV})$\\
\hline 659.28 \qquad\qquad& 4.773 \qquad\qquad& 5.32\qquad\qquad& 250.0\qquad\qquad& 92.4\qquad\qquad& 139.3\\
\hline
\end{tabular}
\end{center}
\end{table}

\section{Mesonic excitations at finite temperature}

Before we study the properties of mesonic excitations at finite
temperature in the PNJL model, the gap equation whose solution
provides the constituent mass of the quark should be given. As
presented in Ref.~\cite{Hansen2007}, such gap equation in the
Hartree approximation reads
\begin{equation}
m=m_{0}+2GT\mathrm{Tr}\sum_{n=-\infty}^{+\infty}\int_{\Lambda}\frac{\mathrm{d}^{3}p}{(2\pi)^{3}}
\frac{-1}{p\!\!\!\slash-m+\gamma^{0}(-iA_{4})},\label{gap}
\end{equation}
where the imaginary time formalism is used and the temporal
component of the four-momentum is discretized, i.e.
$p_{0}=i\omega_{n}$ and $\omega_{n}=(2n+1)\pi T$ is the Matsubara
frequency for a fermion; $m$ is the constituent mass of the quark;
$\mathrm{Tr}$ is the trace which operates over Dirac, flavor, and
color spaces. Here the three-momentum cut-off is employed. After a
sum of the Matsubara frequencies, Eq.~\eqref{gap} can be written as
\begin{equation}
m=m_{0}+2GN_{f}\sum_{c=1}^{N_{c}}\int_{\Lambda}\frac{\mathrm{d}^{3}p}{(2\pi)^{3}}\frac{2m}{E_{p}}
\left\{1-f\left[E_{p}-(-i{A_{4}}_{cc})\right]-f\left[E_{p}+(-i{A_{4}}_{cc})\right]\right\},\label{gap2}
\end{equation}
where $E_{p}=(p^{2}+m^{2})^{1/2}$ and the summation over the color
index can be further written as
\begin{eqnarray}
&&\sum_{c=1}^{N_{c}}f\left[E_{p}-(-i{A_{4}}_{cc})\right]\nonumber \\
& =&\sum_{c=1}^{N_{c}}\frac{1}{e^{\beta E_{p}}e^{i\beta{A_{4}}_{cc}}+1}\nonumber \\
&=&\left[\left(e^{\beta
E_{p}}e^{i\beta{A_{4}}_{22}}+1\right)\left(e^{\beta
E_{p}}e^{i\beta{A_{4}}_{33}}+1\right)+\left(e^{\beta
E_{p}}e^{i\beta{A_{4}}_{11}}+1\right)\left(e^{\beta
E_{p}}e^{i\beta{A_{4}}_{33}}+1\right)\right.\nonumber \\
&&\left.+\left(e^{\beta E_{p}}e^{i\beta{A_{4}}_{11}}\! + \!
1\right)\left(e^{\beta
E_{p}}e^{i\beta{A_{4}}_{22}} \! + \! 1\right)\right]         
\left[\left(e^{\beta E_{p}}e^{i\beta{A_{4}}_{11}} \! + \!
1\right)\left(e^{\beta E_{p}}e^{i\beta{A_{4}}_{22}} \! + \!
1\right)\left(e^{\beta
E_{p}}e^{i\beta{A_{4}}_{33}} \! + \! 1\right)\right]^{-1}\nonumber\\
&=&N_{c}\frac{\bar{\Phi}e^{-\beta E_{p}}+2\Phi e^{-2\beta
E_{p}}+e^{-3\beta E_{p}}}{1+3\bar{\Phi}e^{-\beta E_{p}}+3\Phi
e^{-2\beta E_{p}}+e^{-3\beta
E_{p}}}=N_{c}f_{\Phi}^{+}(E_{p}),\label{distribution1}
\end{eqnarray}
where the distribution function $f_{\Phi}^{+}(E_{p})$ in the PNJL
model has been given in Ref.~\cite{Hansen2007} with another method
and we follow their notations. We can find that, when
$\Phi=\bar{\Phi}=1$, $f_{\Phi}^{+}(E_{p})$ becomes the conventional
Fermi-Dirac distribution function. In the same way, the summation of
the last term in Eq.~\eqref{gap2} is
\begin{eqnarray}
&&\sum_{c=1}^{N_{c}}f\left[E_{p}+(-i{A_{4}}_{cc})\right]\nonumber \\
& =&N_{c}\frac{\Phi e^{-\beta E_{p}}+2\bar{\Phi} e^{-2\beta
E_{p}}+e^{-3\beta E_{p}}}{1+3\Phi e^{-\beta E_{p}}+3\bar{\Phi}
e^{-2\beta E_{p}}+e^{-3\beta
E_{p}}}=N_{c}f_{\Phi}^{-}(E_{p}).\label{distribution2}
\end{eqnarray}
Finally, the gap equation is given by
\begin{equation}
m=m_{0}+2GN_{f}N_{c}\int_{\Lambda}\frac{\mathrm{d}^{3}p}{(2\pi)^{3}}\frac{2m}{E_{p}}
\left[1-f_{\Phi}^{+}(E_{p})-f_{\Phi}^{-}(E_{p})\right].\label{gap3}
\end{equation}
The gap equation in the PNJL model at finite temperature can also be
simply derived from the gap equation at zero temperature, which is
\begin{equation}
m=m_{0}+8GmN_{f}N_{c}iI_{1},\label{gap4}
\end{equation}
where
\begin{equation}
I_{1}=\int\frac{\mathrm{d}^{4}p}{(2\pi)^{4}}\frac{1}{p^{2}-m^{2}}.\label{I1}
\end{equation}
To calculate the integral $I_{1}$ at finite temperature in the PNJL
model, we just need to replace the integral in $p_{0}$ with
$iT\sum_{n}\frac{1}{N_{c}}\sum_{c}$ with $p_{0}=
i\omega_{n}-i{A_{4}}_{cc}$, i.e.
\begin{eqnarray}
I_{1}&=&iT\sum_{n=-\infty}^{+\infty}\frac{1}{N_{c}}\sum_{c=1}^{N_{c}}\int_{\Lambda}\frac{\mathrm{d}^{3}p}{(2\pi)^{3}}
\frac{1}{(i\omega_{n}-i{A_{4}}_{cc})^{2}-{E_{p}}^{2}}\nonumber \\
&
=&-i\int_{\Lambda}\frac{\mathrm{d}^{3}p}{(2\pi)^{3}}\frac{1}{2E_{p}}
\left[1-f_{\Phi}^{+}(E_{p})-f_{\Phi}^{-}(E_{p})\right].
\end{eqnarray}

\begin{figure}[!htb]
\centering
\includegraphics[scale=0.8]{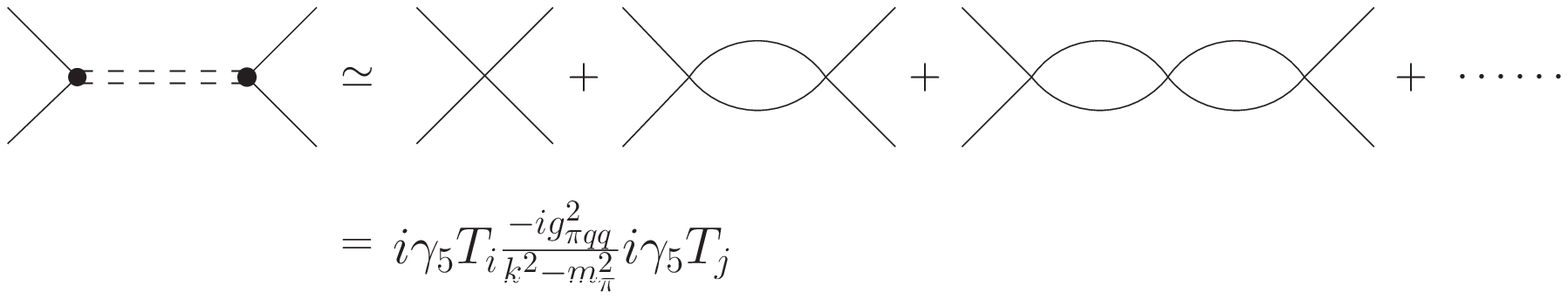}
\caption{Schematic representation of the effective interaction for
the pseudoscalar modes in the RPA, where the double dashed line
represents the effective propagator of $\pi$ mesons and the solid
lines are quark lines; the black dots denote the effective coupling
between $\pi$ meson and quarks. Here, $T_{i}=T_{j}=\tau_{3}$ for
$\pi^{0}$, and $T_{i}=\tau^{\pm}$, $T_{j}=\tau^{\mp}$ for
$\pi^{\pm}$.} \label{f1}
\end{figure}

We have shown that calculations at finite temperatures in the PNJL
model can be simply derived from calculations at zero temperature
above. Therefore, in the following we investigate the mesonic
excitations at finite temperature in the PNJL model starting from
those at zero temperature. We follow the formalism in
Ref.~\cite{Klevansky1992} and the $\pi$ and $\sigma$ mesons
correspond to the pseudoscalar isovector modes and the scalar
isoscalar mode, respectively. For the pseudoscalar modes, defining
the operators
\begin{equation}
\tau^{\pm}=\frac{1}{\sqrt{2}}(\tau_{1}\pm i\tau_{2}),
\end{equation}
we can reexpress the four-fermion term in the pseudoscalar channel
in the Lagrangian in Eq.~\eqref{lagragian} as
\begin{equation}
\left(\bar{\psi}i\gamma_{5}\vec{\tau}\psi\right)^{2}=2\left(\bar{\psi}i\gamma_{5}\tau^{+}\psi\right)
\left(\bar{\psi}i\gamma_{5}\tau^{-}\psi\right)+\left(\bar{\psi}i\gamma_{5}\tau_{3}\psi\right)
\left(\bar{\psi}i\gamma_{5}\tau_{3}\psi\right).
\end{equation}
The effective interaction resulting from the exchange of a $\pi$
meson can be obtained as an infinite sum of loops in the
random-phase approximation (RPA)~\cite{Klevansky1992} and the
leading order terms in $N_{c}$ is shown diagrammatically in
Fig.~\ref{f1}.

Using the symbols in Ref.~\cite{Klevansky1992}, the left hand side
of the equation in Fig.~\ref{f1} can be denoted as $iU_{ij}(k^{2})$.
Summing up all the terms on the right-hand side, we obtain
\begin{equation}
iU_{ij}(k^{2})=i\gamma_{5}T_{i}\frac{2iG}{1-2G\Pi_{ps}(k^{2})}i\gamma_{5}T_{j}.\label{Uij}
\end{equation}
Comparing Eq.~\eqref{Uij} with the equation in Fig.~\ref{f1}, one
can find that the mass of $\pi$ mesons is related to the pole of
Eq.~\eqref{Uij}, which is the solution of the following
equation~\cite{Klevansky1992}
\begin{equation}
1-2G\Pi_{ps}(k^{2})=0.\label{pion_equation}
\end{equation}
Furthermore, the coupling strength between $\pi$ meson and quarks
$g_{\pi qq}$ can be obtained as
\begin{equation}
g_{\pi qq}^{2}=\left[\frac{\partial \Pi_{ps}(k^{2})}{\partial
k^{2}}\right]^{-1}\bigg|_{k^{2}=m_{\pi}^{2}}.\label{pion_coupling}
\end{equation}
Therefore, the information of $\pi$ mesons is included in the
pseudoscalar polarization $\Pi_{ps}(k^{2})$, which reads
\begin{equation}
-i\Pi_{ps}(k^{2})=-\int\frac{\mathrm{d}^{4}p}{(2\pi)^{4}}
\mathrm{Tr}\left[i\gamma_{5}T_{i}iS(k+p)i\gamma_{5}T_{j}iS(p)\right],\label{pion_polarization}
\end{equation}
where $iS(p)=i/(p\!\!\!\slash-m)$ is the propagator of quarks. After
calculating the trace in Eq.~\eqref{pion_polarization}, one has
\begin{eqnarray}
-i\Pi_{ps}(k^{2})&=&4N_{c}N_{f}\int\frac{\mathrm{d}^{4}p}{(2\pi)^{4}}
\frac{1}{p^{2}-m^{2}}-2N_{c}N_{f}k^{2}\int\frac{\mathrm{d}^{4}p}{(2\pi)^{4}}
\frac{1}{(p^{2}-m^{2})[(k+p)^{2}-m^{2}]}\nonumber\\
&=&4N_{c}N_{f}I_{1}-2N_{c}N_{f}k^{2}I(k),\label{pion_polarization2}
\end{eqnarray}
where we have used the function $I_{1}$ given in Eq.~\eqref{I1} and
also defined the function $I(k)$ with the same symbols as used in
Refs.~\cite{Klevansky1992,Schulze1995,Quack1995}, i.e.
\begin{equation}
I(k)=\int\frac{\mathrm{d}^{4}p}{(2\pi)^{4}}
\frac{1}{(p^{2}-m^{2})[(k+p)^{2}-m^{2}]}.\label{Ik}
\end{equation}
Furthermore, we introduce another two functions as done in
Refs.~\cite{Schulze1995,Quack1995}, which will be used in the
following:
\begin{equation}
K(k)=\int\frac{\mathrm{d}^{4}p}{(2\pi)^{4}}
\frac{1}{(p^{2}-m^{2})^{2}[(k+p)^{2}-m^{2}]},\label{Kk}
\end{equation}
\begin{equation}
L(k)=\int\frac{\mathrm{d}^{4}p}{(2\pi)^{4}}
\frac{1}{(p^{2}-m^{2})^{2}[(k+p)^{2}-m^{2}]^{2}}.\label{Lk}
\end{equation}
Then, substituting the expression of the pseudoscalar polarization
in Eq.~\eqref{pion_polarization2} into Eq.~\eqref{pion_equation}, we
have
\begin{equation}
1-8GN_{c}N_{f}iI_{1}+4GN_{c}N_{f}k^{2}iI(k)=0.\label{pion_equation2}
\end{equation}
Upon inserting the gap equation (in Eq.~\eqref{gap4}) into the above
equation, one obtains~\cite{Klevansky1992}
\begin{equation}
\frac{m_{0}}{m}+4GN_{c}N_{f}k^{2}iI(k)=0,\label{pion_equation2}
\end{equation}
whose solution gives the mass of the pseudoscalar mode. The explicit
expression for the coupling between $\pi$ meson and quarks can be
easily obtained upon substituting Eq.~\eqref{pion_polarization2}
into Eq.~\eqref{pion_coupling}, and, in turn, it
reads~\cite{Schulze1995}
\begin{equation}
g_{\pi
qq}^{2}=\frac{i}{N_{c}N_{f}}\frac{1}{I(m_{\pi})+I(0)-m_{\pi}^{2}K(m_{\pi})}.\label{pion_coupling2}
\end{equation}

In the case of finite temperature in the PNJL model, we need to
extend the function $I(k)$ in the same way as taken for the function
$I_{1}$. When the three momentum is vanishing, i.e.
$k=(\omega,\,0)$, $I(\omega,\,0)$ at finite temperature in the PNJL
model is
\begin{equation}
I(\omega,\,0)=-i\int_{\Lambda}\frac{\mathrm{d}^{3}p}{(2\pi)^{3}}
\frac{1}{E_{p}(\omega^{2}-4E_{p}^{2})}\left[1-f_{\Phi}^{+}(E_{p})-f_{\Phi}^{-}(E_{p})\right].\label{Ik2}
\end{equation}
Then Eq.~\eqref{pion_equation2} at finite temperature can be
rewritten as
\begin{equation}
\frac{m_{0}}{m}+4GN_{c}N_{f}m_{\pi}^{2}\int_{\Lambda}\frac{\mathrm{d}^{3}p}{(2\pi)^{3}}
\frac{1}{E_{p}(m_{\pi}^{2}-4E_{p}^{2})}\left[1-f_{\Phi}^{+}(E_{p})-f_{\Phi}^{-}(E_{p})\right]=0.\label{pion_equation3}
\end{equation}
Furthermore, we have
\begin{equation}
K(\omega,\,0)=i\int_{\Lambda}\frac{\mathrm{d}^{3}p}{(2\pi)^{3}}
\frac{\omega^{2}-12E_{p}^{2}}{4E_{p}^{3}(\omega^{2}-4E_{p}^{2})^{2}}
\left[1-f_{\Phi}^{+}(E_{p})-f_{\Phi}^{-}(E_{p})\right],\label{Kk2}
\end{equation}
\begin{equation}
L(\omega,\,0)=i\int_{\Lambda}\frac{\mathrm{d}^{3}p}{(2\pi)^{3}}
\frac{\omega^{2}-20E_{p}^{2}}{2E_{p}^{3}(\omega^{2}-4E_{p}^{2})^{3}}
\left[1-f_{\Phi}^{+}(E_{p})-f_{\Phi}^{-}(E_{p})\right].\label{Lk2}
\end{equation}

In the same way, properties of $\sigma$ mesons can be extracted from
the scalar polarization $\Pi_{s}(k^{2})$, which is
\begin{eqnarray}
-i\Pi_{s}(k^{2})&=&-\int\frac{\mathrm{d}^{4}p}{(2\pi)^{4}}
\mathrm{Tr}\left[iS(k+p)iS(p)\right]\nonumber\\
&=&4N_{c}N_{f}I_{1}-2N_{c}N_{f}(k^{2}-4m^{2})I(k).\label{sigma_polarization}
\end{eqnarray}
Therefore, employing the RPA approximation for the scalar channel in
the same procedure as for the effective interaction in the
pseudoscalar channel, one could determine the mass of $\sigma$ meson
which is the pole of its corresponding effective propagator, i.e.
$1-2G\Pi_{s}(m_{\sigma}^{2})=0$, explicitly given by
\begin{equation}
\frac{m_{0}}{m}+4GN_{c}N_{f}(m_{\sigma}^{2}-4m^{2})iI(m_{\sigma})=0.\label{sigma_equation}
\end{equation}
The relation between the mass of $\sigma$ meson and that of $\pi$
meson could be obtained by comparing Eq.~\eqref{sigma_equation} and
Eq.~\eqref{pion_equation2} as
\begin{equation}
m_{\sigma}^{2}=4m^{2}+m_{\pi}^{2}\frac{I(m_{\pi})}{I(m_{\sigma})},\label{mass_relation}
\end{equation}
which returns to the relation given in Ref.~\cite{Klevansky1992}
when the difference between $I(m_{\pi})$ and $I(m_{\sigma})$ is
neglected.

In the following, we would investigate the pion-decay constant
$f_{\pi}$ at finite temperature in the PNJL model with a starting of
the definition of $f_{\pi}$
\begin{equation}
\langle 0 |
J^{i}_{5\mu}(x)|\pi^{j}\rangle=ik_{\mu}f_{\pi}\delta^{ij} .
\label{fpi_definition}
\end{equation}
Considering the explicit expression of the left hand side in the
PNJL model
\begin{equation}
\langle 0 |
J^{i}_{5\mu}(x)|\pi^{j}\rangle=-\int\frac{\mathrm{d}^{4}p}{(2\pi)^{4}}
\mathrm{Tr}\left[i\gamma_{\mu}\gamma_{5}\frac{\tau^{i}}{2}iS(k+p)ig_{\pi
qq}\gamma_{5}\tau^{j}iS(p)\right] ,
\end{equation}
we arrive at
\begin{equation}
f_{\pi}=-4iN_{c}g_{\pi qq}mI(m_{\pi}).\label{fpi}
\end{equation}
Employing the expression of the pion-quark coupling in
Eq.~\eqref{pion_coupling2} and considering the fact $N_{f} =2$ in
the present case, one has~\cite{Schulze1995}
\begin{equation}
f_{\pi}^{2}=-8iN_{c}m^{2}\frac{I^{2}(m_{\pi})}{I(0)+I(m_{\pi})-m_{\pi}^{2}K(m_{\pi})},\label{fpi2}
\end{equation}
and
\begin{equation}
f_{\pi}^{2}g_{\pi
qq}^{2}=4m^{2}\frac{I^{2}(m_{\pi})}{\left[I(0)+I(m_{\pi})-m_{\pi}^{2}K(m_{\pi})\right]^{2}}\equiv
m^{2}r^{2},\label{GT}
\end{equation}
where we have defined a symbol $r$ as
\begin{equation}
r\equiv
\frac{2I(m_{\pi})}{I(0)+I(m_{\pi})-m_{\pi}^{2}K(m_{\pi})}.\label{r}
\end{equation}
When the temperature is approaching zero, $I(m_{\pi}) \approx I(0)$,
$K(m_{\pi}) \approx 0$, $r\rightarrow 1$.
Eq.~\eqref{GT} returns then to the quark level version of the
Goldberger-Treiman relation at zero
temperature~\cite{Goldberger1958}. As we show below, when the
temperature is near some critical temperature, $r$ deviates from $1$
evidently. Furthermore, from Eq.~\eqref{pion_equation2}, one has the
mass of $\pi$ meson as
\begin{equation}
m_{\pi}^{2}=-\frac{m_{0}}{m}\frac{1}{4GN_{c}N_{f}iI(m_{\pi})}.
\end{equation}
One may also combine this equation with Eq.~\eqref{fpi2}. It gives
consequently
\begin{equation}
m_{\pi}^{2}f_{\pi}^{2}=\frac{m_{0}m}{G}\frac{I(m_{\pi})}{I(0)+I(m_{\pi})-m_{\pi}^{2}K(m_{\pi})}.\label{GM}
\end{equation}
And the constituent mass of the quark $m$ is related with the
condensate of quark by
\begin{eqnarray}
m&=&-2GN_{f}\langle \bar{u}u\rangle+m_{0}\nonumber \\
&=&-2G\langle \bar{\psi}\psi\rangle+m_{0}.\label{condensate}
\end{eqnarray}
Replacing the constituent mass in Eq.~\eqref{GM} with the quark
condensate in Eq.~\eqref{condensate}, we obtain
\begin{equation}
m_{\pi}^{2}f_{\pi}^{2}=-m_{0}\langle \bar{\psi}\psi\rangle
r+\frac{m_{0}^{2}}{2G}r = -m_{0}\langle \bar{\psi}\psi\rangle r
\Big[ 1 + \frac{m_{0}}{2G \vert \langle \bar{\psi}\psi\rangle \vert
} \Big] \, .\label{GM2}
\end{equation}
As mentioned above, at zero temperature $r\rightarrow 1$.
Considering the lowest-order contribution in $m_{0}$, one obtains
then $m_{\pi}^{2}f_{\pi}^{2}\simeq -m_{0}\langle
\bar{\psi}\psi\rangle$, which is the lowest-order approximation to
the Gell-Mann--Oakes--Renner relation~\cite{Gell1968}.

Before our numerical calculations, we need to give the equations to
determine the values of the Polyakov-loop $\Phi$ and its conjugate
$\bar{\Phi}$. In the mean-field approximation or equivalently the
Hartree approximation, the thermodynamical potential density for the
Lagrangian density in Eq.~\eqref{lagragian} is given
as~\cite{Hansen2007}
\begin{eqnarray}
\Omega(\Phi,\bar{\Phi},m,T)&=&\frac{(m_{0}-m)^{2}}{4G}+\mathcal{U}(\Phi,\bar{\Phi},T)
-2N_{f}N_{c}\int_{\Lambda}\frac{\mathrm{d}^{3}p}{(2\pi)^{3}}E_{p}\nonumber \\
&&-2N_{f}T\int_{\Lambda}\frac{\mathrm{d}^{3}p}{(2\pi)^{3}}\left[
\ln\left(1+N_{c}\bar{\Phi}e^{-\beta E_{p}}+N_{c}\Phi e^{-2\beta
E_{p}}+e^{-3\beta E_{p}}\right)\right.\nonumber \\
&&\left.+\ln\left(1+N_{c}\Phi e^{-\beta E_{p}}+N_{c}\bar{\Phi}
e^{-2\beta E_{p}}+e^{-3\beta E_{p}}\right)\right].\label{potential}
\end{eqnarray}
Minimizing this thermodynamical potential with respect to $\Phi$ and
$\bar{\Phi}$, we obtain equations
\begin{equation}
\frac{\partial \Omega}{\partial \Phi}=0,\qquad \frac{\partial
\Omega}{\partial \bar{\Phi}}=0.\label{Eq_Phi}
\end{equation}
In the absence of chemical potential, these two equations are
identical, and so $\Phi=\bar{\Phi}$~\cite{Ratti2006a}. In the same
way, Minimizing the thermodynamical potential in
Eq.~\eqref{potential} with respect to the value of the constituent
quark mass $m$, the gap equation in Eq.~\eqref{gap3} can also be
obtained.

In the following, we present our numerical results. First of all, we
give our calculated values for several characteristic temperatures.
These characteristic temperatures include the pseudo-transition
temperature for chiral crossover, $T_{\chi}$, corresponding to the
maximum of
$-\mathrm{d}m/\mathrm{d}T$~\cite{Ratti2006a,Fu2008,Hansen2007}, the
pseudo-transition temperature for deconfinement crossover, $T_{P}$,
corresponding to the maximum of $\mathrm{d}\Phi/\mathrm{d}T$, the
Mott temperature $T_{M}$ for $\pi$ meson, defined by
\begin{equation}
m_{\pi}(T_{M})=2m(T_{M}),
\end{equation}
meaning that the pion can dissociate into a constituent quark and an
antiquark above the Mott temperature, and the dissociation
temperature for $\sigma$ meson $T_{d}^{\sigma}$~\cite{Quack1995},
defined by
\begin{equation}
m_{\sigma}(T_{d}^{\sigma})=2m_{\pi}(T_{d}^{\sigma}).
\end{equation}
These characteristic temperatures except for $T_{P}$, can also serve
in the conventional NJL model. Numerical results for these
characteristic temperatures in the PNJL model with two Polyakov-loop
effective potentials are shown in Table~\ref{temperatues}. Here, for
comparison we also list the results in the conventional NJL model.

\begin{table}[htb]
\begin{center}
\caption{Several critical temperatures in the conventional NJL model
and the PNJL model with two Polyakov-loop effective potentials
($T_{0}=270\,\mathrm{MeV}$ is chosen for these two effective
potentials).} \label{temperatues}
\begin{tabular}{ccccc}
\hline \hline \vspace{0.1cm}
&$T_{\chi}\,(\mathrm{MeV})$\qquad\qquad&$T_{P}\,(\mathrm{MeV})$\qquad\qquad&$T_{M}\,(\mathrm{MeV})$\qquad\qquad&
$T_{d}^{\sigma}\,(\mathrm{MeV})$\\
\hline PNJL ($\mathcal{U}_{\mathrm{pol}}$)\qquad\qquad& 253.2 \qquad\qquad& 245.4\qquad\qquad& 264.6\qquad\qquad& 253.0\\
\hline PNJL ($\mathcal{U}_{\mathrm{imp}}$)\qquad\qquad& 245.0 \qquad\qquad& 232.0\qquad\qquad& 259.6\qquad\qquad& 246.3\\
\hline NJL \qquad\qquad& 184.4 \qquad\qquad&--- \qquad\qquad& 201.2\qquad\qquad& 181.9\\
\hline
\end{tabular}
\end{center}
\end{table}

\begin{figure}
\centering
\includegraphics[scale=0.75,angle=0]{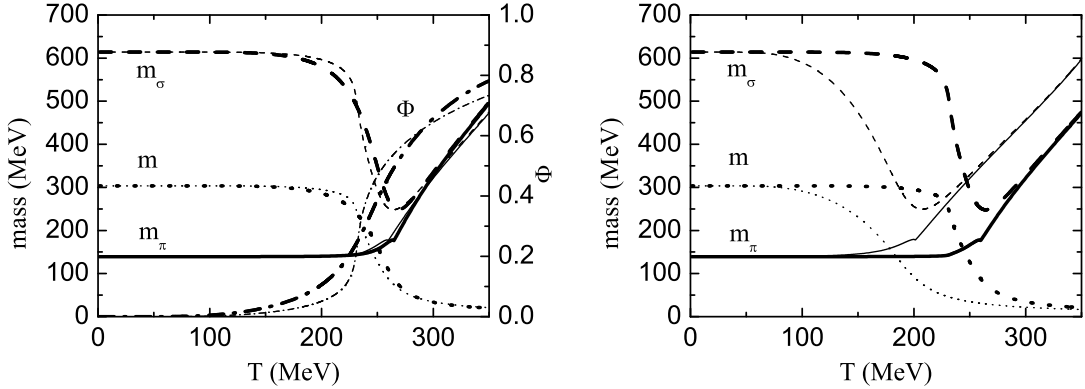}
\caption{\label{f2} Left panel: calculated masses of $\pi$ meson,
$\sigma$ meson, constituent quark and the Polyakov-loop as functions
of the temperature. Here, thick curves and thin curves correspond to
the results with the polynomial Polyakov-loop effective potential
$\mathcal{U}_{\mathrm{pol}}$, the improved effective potential
$\mathcal{U}_{\mathrm{imp}}$, respectively. Right panel: masses of
$\pi$, $\sigma$, and $m$ as functions of the temperature in the PNJL
model with $\mathcal{U}_{\mathrm{imp}}$ (thick curves) and in the
conventional NJL model (thin curves).}
\end{figure}

In Fig.~\ref{f2} we illustrate our calculated results of the masses
of $\pi$ and $\sigma$ mesons, the mass of constituent quark, and the
Polyakov-loop as functions of the temperature. Fig.~\ref{f2} shows
evidently that at a temperature not very high, the masses of the
constituent quark, the pion and the $\sigma$ mesons maintain the
same as the corresponding one at zero temperature. As the
temperature is around the critical one, these masses vary abruptly.
And further, if the temperature is very high, the masses of $\pi$
and $\sigma$ mesons become degenerate, which indicates that the
chiral symmetry is restored at high temperature. Such a feature is
consistent with that given in the framework of Bethe-Salpeter
equation combining with the Dyson-Schwinger equations (see for
example Ref.~\cite{Maris2001}). We also find that two different
Polyakov-loop effective potentials do not result in qualitative
differences but only slightly quantitative deviations as the left
panel of Fig.~\ref{f2} shows. Furthermore, looking through the right
panel of Fig.~\ref{f2}, we can notice that the chiral phase
transition occurs at relatively lower temperature in the
conventional NJL model.

\begin{figure}
\centering
\includegraphics[scale=0.75,angle=0]{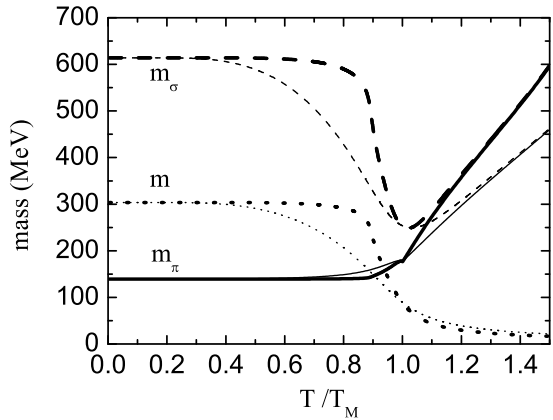}
\caption{\label{f3} Calculated masses of $\pi$ meson, $\sigma$
meson, and constituent quark as functions of the temperature in unit
of Mott temperature $T_{M}$ in the PNJL model with
$\mathcal{U}_{\mathrm{imp}}$ (thick lines) and in the conventional
NJL model (thin lines).}
\end{figure}

In order to compare the obtained results in the PNJL model with
those in the conventional NJL model more conveniently, we scale the
temperature in unit of Mott temperature $T_{M}$ and re-display the
results in Fig.~\ref{f3}. One can recognize that, in the PNJL model,
only when the temperature is very near the phase transition
temperature, masses of mesons and constituent quark begin to deviate
from their values at zero temperature obviously. While in the
conventional NJL model, these masses begin to deviate from their
zero-temperature values at much lower temperature, about
$0.4\,T_{M}$. This phenomenon can be attributed to the fact that, in
the low temperature, chiral symmetry is broken and the quark and
antiquark are in the confined hadronic phase in the PNJL model,
contributions from thermal excitations of one and two quarks or
antiquarks are suppressed as the distribution functions in
Eq.~\eqref{distribution1} and Eq.~\eqref{distribution2} show when
the Polyakov-loop $\Phi$ approaches zero. This is a manifestation of
color confinement on the level of statistics. While in the
conventional NJL model, due to the lack of the appearance of color
confinement, contributions from one and two quarks or antiquarks
become significant even at low temperature, which results in the
phenomenon mentioned above.

\begin{figure}
\centering
\includegraphics[scale=0.75,angle=0]{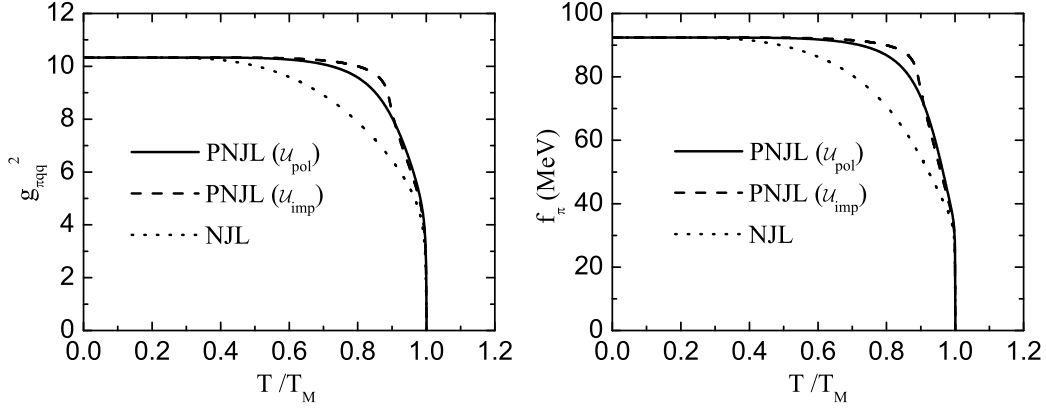}
\caption{\label{f4} Left panel: calculated results of the square of
the pion-quark coupling strength $g_{\pi qq}^{2}$ (in
Eq.~\eqref{pion_coupling2}) as a function of the temperature in unit
of Mott temperature $T_{M}$ in the PNJL and the conventional NJL
model. Right panel: calculated results of the pion-decay constant
$f_{\pi}$ (in Eq.~\eqref{fpi}) as a function of the temperature in
unit of Mott temperature.}
\end{figure}

In Fig.~\ref{f4} we show the square of the pion-quark coupling
strength $g_{\pi qq}^{2}$ and the pion-decay constant $f_{\pi}$ as
functions of the temperature in unit of Mott temperature in the PNJL
model with polynomial and improved effective potentials and in the
conventional NJL model. As Eq.~\eqref{Kk2} shows, when temperature
approaches the Mott temperature $T_{M}$ from below, i.e. when the
mass of $\pi$ meson is about twice mass of the constituent quark,
$iK(m_{\pi})\rightarrow \infty$. Therefore, the pion-quark coupling
strength and pion-decay constant vanish at $T_{M}$, as
Eq.~\eqref{pion_coupling2} and Eq.~\eqref{fpi2} show. Furthermore,
One can also find in Fig.~\ref{f4} that, in the PNJL model, $g_{\pi
qq}^{2}$ and $f_{\pi}$ almost keep invariant with the increase of
the temperature when the temperature is not high and these two
quantities decrease rapidly only when the temperature is above
$0.8\,T_{M}$. While in the conventional NJL model these two
quantities begin to decrease at about $0.4\,T_{M}$. This behavior is
also due to the lack of the color confinement in the conventional
NJL model as the same as the behavior of masses of mesons and
constituent quark as functions of temperature shown in
Fig.~\ref{f3}.

\begin{figure}
\centering
\includegraphics[scale=0.75,angle=0]{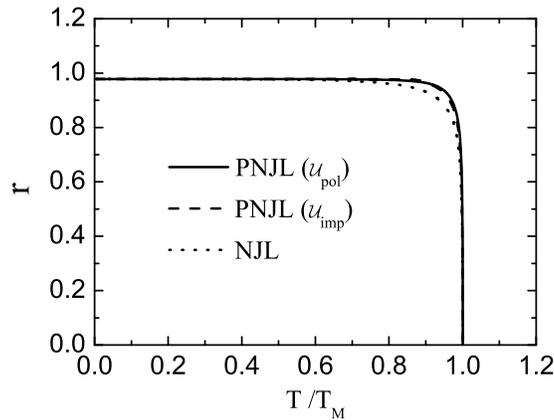}
\caption{\label{f5} Calculated factor $r$ defined in Eq.~\eqref{r}
as a function of the temperature in unit of Mott temperature in the
PNJL and the conventional NJL model.}
\end{figure}

We have shown above that the Goldberger-Treiman relation and
Gell-Mann--Oakes-Renner relation at finite temperature are different
from those at zero temperature in that a factor $r$ defined in
Eq.~\eqref{r} is introduced. The calculated behavior of $r$ as
function of temperature is displayed in Fig.~\ref{f5} and one can
find that, when the temperature is below $0.9\,T_{M}$, $r$ is almost
a constant very near $1$, which indicates that these two important
relations at vacuum still serve well in the large region of
temperature $0\sim 0.9\,T_{M}$. However, when the temperature is
above $0.9\,T_{M}$, $r$ decreases very rapidly and vanishes at
$T=T_{M}$. In the region of the temperature $0.9\,T_{M} \sim T_{M}$,
the Goldberger-Treiman relation and Gell-Mann--Oakes-Renner relation
at vacuum should be extended to Eq.~\eqref{GT}, Eq.~\eqref{GM2},
respectively.

\section{$\pi$--$\pi$ scattering lengths}

\begin{figure}[!htb]
\centering
\includegraphics[scale=0.8]{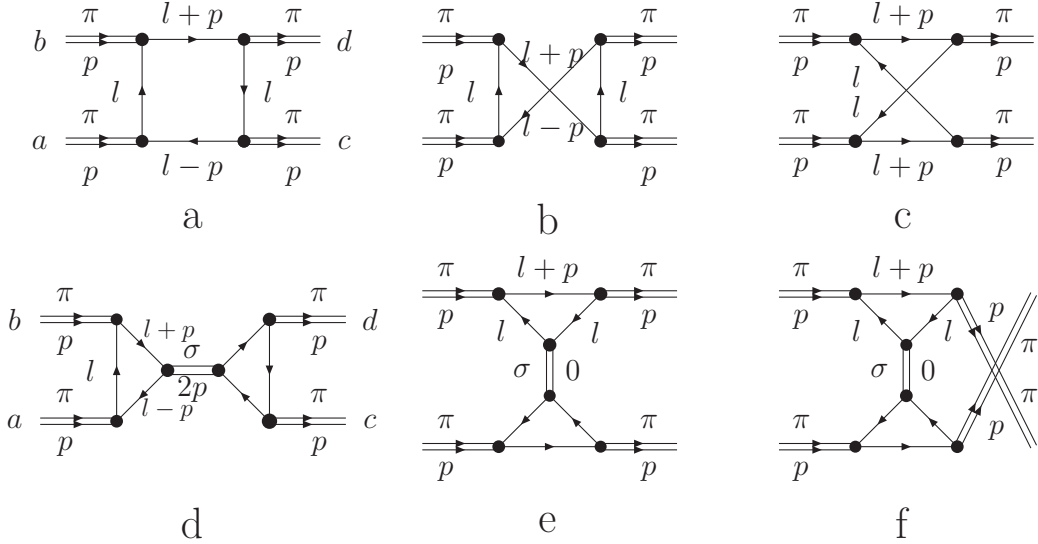}
\caption{Feynman diagrams contributing to the $s$-wave $\pi$--$\pi$
scattering (see also Ref.~\cite{Bernard1991,Schulze1995}). Here the
external momenta for pions are chosen to be the special case in
Eq.~\eqref{momenta}.} \label{f6}
\end{figure}

The formalism of $s$-wave $\pi$--$\pi$ scattering lengths at zero
temperature in the conventional NJL model has been established in
Refs.~\cite{Bernard1991,Bernard1992,Schulze1995}, and it has been
extended to the case at finite temperature by Quack et
al.~\cite{Quack1995}. In this work we follow the notation and
calculation given in Ref.~\cite{Schulze1995}. The invariant
amplitude of $\pi$--$\pi$ scattering has the form:
\begin{equation}
\langle\,
cp_{c};dp_{d}|i\mathcal{M}|ap_{a};bp_{b}\,\rangle=iA(s,t,u)\delta_{ab}\delta_{cd}
+iB(s,t,u)\delta_{ac}\delta_{bd}+iC(s,t,u)\delta_{ad}\delta_{bc},\label{total_am}
\end{equation}
where $a$, $b$, $c$, and $d$ are the isospin labels, and $s$, $t$
and $u$ are the Mandelstam variables, $s=(p_{a}+p_{b})^{2}$,
$t=(p_{a}-p_{c})^{2}$ and $u=(p_{a}-p_{d})^{2}$. The amplitude of
definite total isospin $I$, defined by $A_{I}$, can be projected
out, given by Ref.~\cite{Schulze1995}
\begin{equation}
A_{0}=3A+B+C, \quad A_{1}=B-C,\; \mathrm{and} \quad
A_{2}=B+C.\label{A_I}
\end{equation}
When the scattering is at the kinematic threshold, we obtain the
scattering lengths, i.e.
\begin{equation}
a_{I}=\frac{1}{32\pi}A_{I}(s=4m_{\pi}^{2}, t=0, u=0).\label{length}
\end{equation}
For simplicity, the pion momenta can be chosen as
\begin{equation}
p_{a}=p_{b}=p_{c}=p_{d}=p,\; \mathrm{and}\quad
p^{2}=m_{\pi}^{2},\label{momenta}
\end{equation}
which can be verified to fulfill the threshold condition in
Eq.~\eqref{length}. To lowest order in $1/N_{c}$, there are two
types of Feynman diagrams contributing to the $s$-wave $\pi$--$\pi$
scattering~\cite{Bernard1991,Schulze1995}, i.e.the box diagram and
the $\sigma$-propagation diagram. Here we also present them in
Fig.~\ref{f6}. The three diagrams in the first row of Fig.~\ref{f6}
are the box diagrams and the ones in the second row are the
$\sigma$-propagation diagrams.

Following the calculation of Ref.~\cite{Schulze1995}, we obtain
respective amplitude for each diagram in Fig.~\ref{f6} as
\begin{eqnarray}
i\mathcal{M}_{a}&=&(\delta_{ab}\delta_{cd}+\delta_{ac}\delta_{bd}-\delta_{ad}\delta_{bc})
(-4N_{c}N_{f}g_{\pi qq}^{4})[I(0)+I(p)-p^{2}K(p)]\nonumber \\
&\equiv&(\delta_{ab}\delta_{cd}+\delta_{ac}\delta_{bd}-\delta_{ad}\delta_{bc})iT_{a},\label{a}\\
i\mathcal{M}_{b}&=&(\delta_{ab}\delta_{cd}-\delta_{ac}\delta_{bd}+\delta_{ad}\delta_{bc})
(-4N_{c}N_{f}g_{\pi qq}^{4})[I(0)+I(p)-p^{2}K(p)]\nonumber \\
&\equiv&(\delta_{ab}\delta_{cd}-\delta_{ac}\delta_{bd}+\delta_{ad}\delta_{bc})iT_{b},\\
i\mathcal{M}_{c}&=&(-\delta_{ab}\delta_{cd}+\delta_{ac}\delta_{bd}+\delta_{ad}\delta_{bc})
(-8N_{c}N_{f}g_{\pi qq}^{4})\left[I(0)+\frac{p^{4}}{2}L(p)-2p^{2}K(p)\right]\nonumber \\
&\equiv&(-\delta_{ab}\delta_{cd}+\delta_{ac}\delta_{bd}+\delta_{ad}\delta_{bc})iT_{c},
\end{eqnarray}
\begin{eqnarray}
i\mathcal{M}_{d}&=&\delta_{ab}\delta_{cd}(8N_{c}N_{f}g_{\pi
qq}^{4})\frac{I^{2}(p)}{\left(1-\frac{p^{2}}{m^{2}}\right)I(2p)+\frac{m_{\pi}^{2}}{4m^{2}}I(m_{\pi})}\nonumber \\
&\equiv&\delta_{ab}\delta_{cd}iT_{d},\\
i\mathcal{M}_{e}&=&\delta_{ac}\delta_{bd}(8N_{c}N_{f}g_{\pi
qq}^{4})\frac{[I(0)-p^{2}K(p)]^{2}}{I(0)+\frac{m_{\pi}^{2}}{4m^{2}}I(m_{\pi})}\nonumber \\
&\equiv&\delta_{ac}\delta_{bd}iT_{e},\\
i\mathcal{M}_{f}&=&\delta_{ad}\delta_{bc}(8N_{c}N_{f}g_{\pi
qq}^{4})\frac{[I(0)-p^{2}K(p)]^{2}}{I(0)+\frac{m_{\pi}^{2}}{4m^{2}}I(m_{\pi})}\nonumber \\
&\equiv&\delta_{ad}\delta_{bc}iT_{f}.\label{f}
\end{eqnarray}
From above equations, one can notice $T_{b}=T_{a}$, and
$T_{f}=T_{e}$. Substituting Eqs.~\eqref{a}---~\eqref{f} into
Eq.~\eqref{total_am}, one obtains
\begin{equation}
A=2T_{a}-T_{c}+T_{d}, \qquad B=C=T_{c}+T_{e}.
\end{equation}
Therefore, employing Eq.~\eqref{A_I} we have
\begin{eqnarray}
A_{0}&=&6T_{a}-T_{c}+3T_{d}+2T_{e},\nonumber \\
A_{1}&=&0,\nonumber \\
A_{2}&=&2(T_{c}+T_{e}).\label{A_I2}
\end{eqnarray}

\begin{figure}[!htb]
\centering
\includegraphics[scale=1.1]{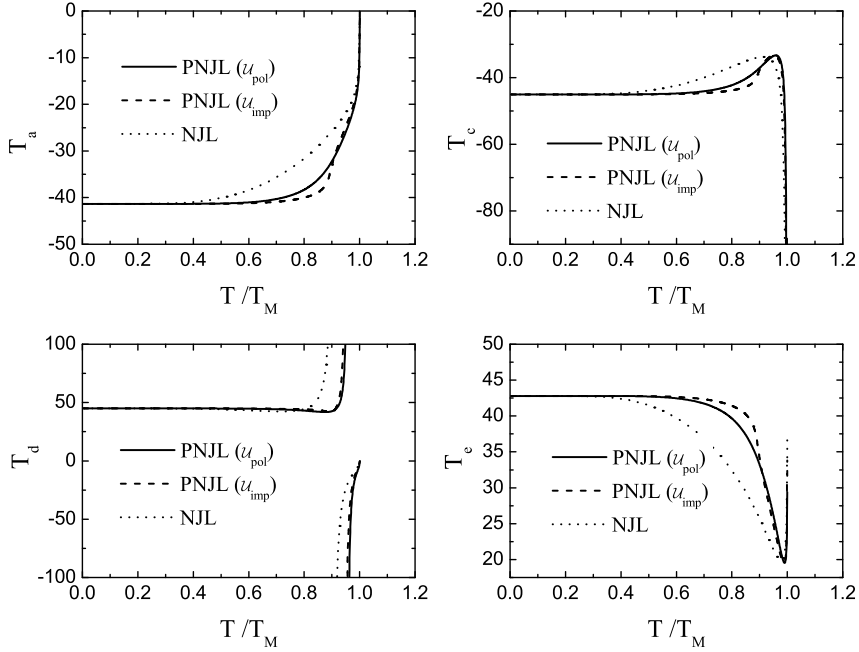}
\caption{Calculated scattering amplitudes $T_{a}$, $T_{c}$, $T_{d}$,
and $T_{e}$ as functions of the temperature in unit of Mott
temperature in the PNJL and the conventional NJL models.} \label{f7}
\end{figure}

In Fig.~\ref{f7} we present our calculated results of the scattering
amplitudes $T_{a}$, $T_{c}$, $T_{d}$, and $T_{e}$ as functions of
the temperature in unit of $T_{M}$ in the PNJL model with polynomial
Polyakov-loop effective potential and improved effective potential
and in the conventional NJL model. The results of the conventional
NJL model in our work are roughly consistent with those given in
Ref.~\cite{Quack1995}. However, there exists a difference which
reads that our present calculation indicates that the scattering
amplitude $T_{a}$ approaches zero at the Mott temperature $T_{M}$,
the calculation in Ref.~\cite{Quack1995} gives that $T_{a}$ is
divergent at $T_{M}$. Recalling the analysis above, we would
emphasized that, when the temperature approaches to $T_{M}$,
$iK(m_{\pi})$ in Eq.~\eqref{Kk2} and $iL(m_{\pi})$ in
Eq.~\eqref{Lk2} are divergent and the degree of divergence of
$iL(m_{\pi})$ is higher than that of $iK(m_{\pi})$. Substituting the
expression of the coupling between $\pi$ meson and quarks $g_{\pi
qq}^{2}$ in Eq.~\eqref{pion_coupling2} into Eq.~\eqref{a}, we find
$T_{a}\propto 1/(-iK(m_{\pi}))$ when the temperature is near the
Mott temperature, therefore, $T_{a}$ approaches zero at $T_{M}$.
In the same way, we find that $T_{d}$ approaches zero, $T_{c}$ is
divergent, and $T_{e}$ approaches a finite value at the Mott
temperature. Furthermore, when the temperature is equal to the
dissociation temperature of $\sigma$ meson, i.e. $T=T_{d}^{\sigma}$,
we have $m_{\sigma}=2m_{\pi}$, which results in that the $\sigma$
propagator in diagram d of Fig.~\ref{f6} and also the amplitude
$T_{d}$ (see Fig.~\ref{f7}) are divergent at $T_{d}^{\sigma}$.
Comparing the results of the PNJL model with those of the
conventional NJL model, one can also recognize the similar behavior
as obtained previously, which reads that the $T$-matrix amplitudes
calculated in the PNJL model deviate from their values at zero
temperature only when the temperature is near the critical
temperature, while the deviation occurs much earlier in the
conventional NJL model. Taking amplitude $T_{d}$ for example, since
the mass of $\sigma$ meson decreases with the increase of the
temperature much earlier in the conventional NJL model than that in
the PNJL model, as Fig.~\ref{f3} shows, we expect that the
divergence in $T_{d}$ also occurs earlier in the conventional NJL
model, which is verified in Fig.~\ref{f7}. In addition, we find that
the value of $T_{d}^{\sigma}/T_{M}$ calculated in the conventional
NJL model is about $0.90$, smaller than $0.96$ in the PNJL model
with polynomial effective potential, and $0.95$ in the PNJL model
with improved effective potential.

\begin{figure}[!htb]
\centering
\includegraphics[scale=0.7]{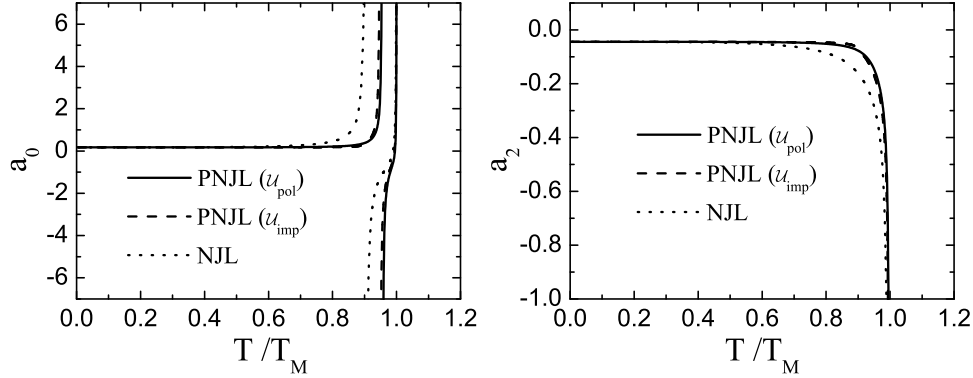}
\caption{Calculated $s$-wave $\pi$--$\pi$ scattering lengths $a_{0}$
and $a_{2}$ as functions of the temperature in unit of Mott
temperature in the PNJL and the conventional NJL models.} \label{f8}
\end{figure}

In Fig.~\ref{f8}, we show the calculated results of the $s$-wave
$\pi$--$\pi$ scattering lengths $a_{0}$ and $a_{2}$ as functions of
the temperature in unit of $T_{M}$ in the PNJL model with two
Polyakov-loop effective potentials and in the conventional NJL
model. Since $a_{0}$ and $a_{2}$ contain contribution from the
$T$-matrix amplitude $T_{c}$ as Eq.~\eqref{A_I2} shows, they are
both divergent at $T=T_{M}$. Furthermore, $a_{0}$ also contains
$T_{d}$, so it diverges at $T=T_{d}^{\sigma}$ as well. At zero
temperature, we have $a_{0}=0.173$ and $a_{2}=-0.045$, which are
consistent with the Weinberg values $(a_{0})^{W}=7m_{\pi}^{2}/(32\pi
f_{\pi}^{2})=0.158$ and $(a_{2})^{W}=-2m_{\pi}^{2}/(32\pi
f_{\pi}^{2})=-0.045$~\cite{Quack1995,Bernard1991}. On the
experimental side, the Geneva-Saclay collaboration provided the
often quoted values $a_{0}=0.26\pm 0.05$ and $a_{2}=-0.028\pm
0.012$~\cite{Froggatt1977,Nagels1979}, and recent years experiment
E865 at Brookhaven National Laboratory, USA, has given new values
$a_{0}=0.203\pm 0.033\pm 0.004$ and $a_{2}=-0.055\pm 0.023 \pm
0.003$~\cite{Truol2000}, and also $a_{0}=0.216\pm 0.013\pm 0.004\pm
0.005$~\cite{Pislak2001}. The scattering lengths $a_{0}$ and $a_{2}$
are almost independent of the temperature until the temperature is
increased to $0.9T_{M}$ in the PNJL model, and after that they vary
drastically with temperature. While in the conventional NJL model
$a_{0}$ and $a_{2}$ begin to vary with temperature at about
$0.6T_{M}$ and the temperature at which $a_{0}$ diverges due to the
$\sigma$-meson dissociation is also lower in the conventional NJL
model.
From the above analysis, one can recognize that the physical meaning
of the divergence of the $s$-wave $\pi$--$\pi$ scattering lengths
$a_{0}$ and $a_{2}$ at the Mott temperature $T_{M}$ for $\pi$ meson
is clear and consistent with that shown in Ref.~\cite{Quack1995}. At
$T=T_{M}$, $\pi$ meson can dissociate into a constituent quark and
an antiquark, the $\pi$ meson is then unbound and its radius becomes
infinite. Mathematically, the geometrical size of the pion meson can
be described by its charge radius $r$ which is related to $f_{\pi}$
through $f_{\pi}^{2}\propto 1/\langle
r^{2}\rangle$~\cite{Quack1995}. Employing the Weinberg relations
cited above, we have the relation between the scattering lengths and
the charge radius of $\pi$ meson as $|a|\propto \langle
r^{2}\rangle$, which clearly indicates that the divergence of
$\pi$--$\pi$ scattering lengths at the pion Mott temperature is
closely related with the melting of the pion meson. The relation
between the divergence of the $s$-wave $\pi$--$\pi$ scattering
lengths and the delocalization of the pion meson at $T_{M}$ is then
confirmed not only in the conventional NJL model but also in the
PNJL model. As for the divergence of the $s$-wave $\pi$--$\pi$
scattering length $a_{0}$ at the dissociation temperature for
$\sigma$ meson $T_{d}^{\sigma}$, we should note that this divergence
corresponds to the situation that the propagator for $\sigma$ meson
displayed in the part d of Fig.~\ref{f6} is on shell, which means
that the $\pi$--$\pi$ scattering in the s channel couples resonantly
with the $\sigma$ meson field.
The divergence (from positive infinite to negative infinite) of the
$a_{0}$ and the mass relation $m_{\sigma} = 2 m_{\pi}$ suggest that,
in general point of view, a very loosely bound state may appear at
the dissociation temperature $T_{d}^{\sigma}$.
However, detailed investigation is required to clarify its
mechanism.

\section{Summary}

In summary, we have studied the mesonic excitations at finite
temperature in the two flavor PNJL model. The masses of $\pi$ meson
and $\sigma$ meson, pion-decay constant, and the pion-quark coupling
strength at finite temperature are calculated in the PNJL model with
two forms of Polyakov-loop effective potential. Their variation
behaviors with temperature, especially when the temperature takes a
value near the critical one, are investigated in details. We find
that the results calculated in the PNJL model are almost independent
of the choice of the Polyakov-loop effective potential. We also
compare our calculated results in the PNJL model with those in the
conventional NJL model. We find that, since in the PNJL model, the
Polyakov-loop which is coupled with quarks suppresses the unwanted
degrees of freedom below the critical temperature, all quantities
describing the properties of mesons deviate from their values at
zero temperature only when the temperature is very near the critical
temperature, for example the Mott temperature, and they vary with
temperature rapidly in a very narrow region near the critical one.
While in the conventional NJL model, these quantities begin to vary
with temperature much earlier. Therefore, we conclude that the
effect of color confinement screens the effect of temperature below
the critical temperature. Furthermore, we have investigated the
Goldberger-Treiman relation and the Gell-Mann--Oakes--Renner
relation at finite temperature in the PNJL model, and we find that
when the temperature is below about $0.9T_{M}$, where $T_{M}$ is the
Mott temperature for $\pi$ meson, these two important relations are
hardly changed by the effect of temperature. However, the
Goldberger-Treiman relation and the Gell-Mann--Oakes--Renner
relation should be corrected once the temperature is in the region
of $0.9T_{M}\sim T_{M}$.

In this work, we have also investigated the $s$-wave $\pi$--$\pi$
scattering lengths at finite temperature in the PNJL model. The
obtained results in the PNJL model are also compared with those in
the conventional NJL model. We find that scattering length $a_{0}$
is divergent at Mott temperature for $\pi$ meson, $T_{M}$, and at
dissociation temperature for $\sigma$ meson, $T_{d}^{\sigma}$, and
scattering length $a_{2}$ is divergent at $T_{M}$, which are
consistent with the results in the conventional NJL model. Due to
the effect of color confinement, the dissociation temperature for
$\sigma$ meson $T_{d}^{\sigma}$ calculated in unit of $T_{M}$ in the
PNJL model is relatively larger than that given in the conventional
NJL model. In the same way, the influence of the temperature on the
scattering lengths $a_{0}$ and $a_{2}$ below the critical
temperature is suppressed by the color confinement in the PNJL
model.
In addition, the characteristic of the scattering amplitude $T_{a}$
at the Mott temperature calculated in the PNJL model is different
from that given previously in the conventional NJL model.

\section*{Acknowledgements}
This work was supported by the National Natural Science Foundation
of China under contract No. 10425521 and No. 10675007, and the Major
State Basic Research Development Program under contract No.
G2007CB815000.
Helpful discussions with Dr. Craig D. Roberts at Argonne National
Laboratory, USA, and Prof. Chuan Liu, Prof. Han-qing Zheng and Dr.
Lei Chang are acknowledged with great thanks.

\end{document}